\documentclass[11pt]{article}
\usepackage{amsmath,amssymb,epsfig,cite,color}
\usepackage{authblk}
\advance\hoffset by -1.1truecm
\advance\voffset by -1.0truecm
\newcommand{\be}{\begin{equation}}
\newcommand{\ee}{\end{equation}}
\newcommand{\bea}{\begin{eqnarray}}
\newcommand{\eea}{\end{eqnarray}}

\numberwithin{equation}{section}

\newcounter{appendice}

 \title{
\vspace{2.2cm} \bf{QCD Breaks Lorentz Invariance and Colour
}}

\author{A.~P.~Balachandran \thanks{balachandran38@gmail.com}}
\affil{\small Physics Department, Syracuse University, Syracuse, New York 13244-1130, U.S.A.}
\date{}

\begin{document}

\maketitle

\begin{abstract}
In a previous work~\cite{ref1}, we have argued that the algebra of non-abelian superselection
rules is spontaneously broken to
its maximal abelian subalgebra, that is, the algebra generated by its completing commuting set (the two Casimirs
and a basis of its Cartan subalgebra). In this paper, alternative arguments confirming these results are
presented. In addition, Lorentz invariance is shown to be broken in QCD, just as it is in QED.
The experimental consequences of these results include fuzzy mass and spin shells of coloured particles like quarks,
and decay life times which depend on the frame of observation~\cite{ref4,ref5,ref7}.

In a paper under preparation, these results are extended to the ADM Poincar\'e group and the local Lorentz
group of frames. The renormalisation of   the ADM energy by infrared gravitons is also studied and estimated.
\end{abstract}

\section{Introduction}

\label{section1}

Quantum field theory (QFT) is defined by the algebra $\mathcal{A}$ of local observables and an irreducible
representation (IRR) $\pi$ of $\mathcal{A}$ on a Hilbert space $\mathcal{H}$. In general, there are many inequivalent
IRR's $\pi_0,\pi_1,\ldots$ of $\mathcal{A}$ defining its superselection sectors.

For example, in QED, $\pi_0$ can be the sector with total charge $q_0=0$, while $\pi_n$ can be the sector
with total charge $q_n$. No observation can mix these sectors.

In QED, the charge operator $Q$ generates the abelian $U(1)$ group. In QCD, the $U(1)$ is replaced that the
non-abelian $SU(3)$ of colour. Superficially, the superselection algebra seems to be the group algebra
$\mathbb{C} SU(3)$. We have previously argued~\cite{ref1} that in reality it is a maximal abelian
subalgebra of $\mathbb{C} SU(3)$ generated by a complete commuting set (CCS).
(See in this connection the Kadison-Singer conjecture and its recent proof~\cite{ref2}.)
We can assume  the CCS to be generated
by the two Casimir operators and the operators on $\mathcal{H}$ representing  the $\lambda_3$ and $\lambda_8$
of  the Gell-Mann  matrices. The remaining independent generators of $\mathbb{C} SU(3)$  are spontaneously broken.

There exist elegant proofs of Roepstorff~\cite{ref3},  Buchholz et. al.~\cite{ref4} and especially
 Fr\"ohlich et. al.~\cite{ref5} that infrared effects in QED break Lorentz invariance. We adapt these arguments here
to show that generic elements of $SU(3)$ map an IRR  $\pi$ of $\mathcal{A}$ to a distinct IRR $\pi^\prime\neq \pi$
of $\mathcal{A}$.
In Higgs theories, this phenomenon is well-known: when the Higgs field $\varphi$ breaks $SO(3)$ to $SO(2)$ say,
as in the  't Hooft-Polyakov model~\cite{ref6}, $SO(3)$ transformations which change the direction of $\varphi$
at spatial infinity cannot be represented as unitary operators on $\mathcal{H}$. In the same way, here,
any operator which disturbs the eigenvalues of the CCS is spontaneously broken.

In section \ref{section2}, we review the QED result on Lorentz violation.  This is then generalised to QCD in sections
\ref{section3} and \ref{section4}.

The results of this paper can be adapted to any non-abelian gauge group.

\section{The Case of QED}

\label{section2}

QED is classified by a continuous family of superselection sectors.

The first is its classification
by the charge $q_n$. In the representation $\pi_n$ of $\mathcal{A}$, the charge operator $Q$ has the eigenvalue $q_n$:
\begin{eqnarray}
\pi_n: \quad Q|n, P,\cdot\rangle=q_n|n,P,\cdot\rangle
\end{eqnarray}
where $P=(P^\mu)$ denotes the total momenta. 
Local observables cannot change $q_n$.

Then, there are the sectors with ``in" state vectors ~\cite{ref7}
\begin{eqnarray}
\label{reprA}
&& e^{q_n\int d^3x [A_i^-(x)\omega_i^+(x)-A_i^+(x)\omega_i^-(x)]}|n, P,\cdot\rangle:=|n, P, \omega,\cdot\rangle, \\
&& |n, P, 0,\cdot\rangle\equiv|n, P, \cdot\rangle, \quad\quad q_n\neq 0
\end{eqnarray}
created by the infrared photons.  Here $A_i^\pm$ are the positive and negative frequency parts of the electromagnetic potential in the Coulomb gauge, and the functions $\omega_i^+$, $\omega_i^- = \bar{\omega}_i^+$  are transverse:
\begin{eqnarray}
\partial_i\omega_i^\pm(x)=0,
\end{eqnarray}

Also they do not vanish fast as we approach infinity:
\begin{eqnarray}
\lim_{r\rightarrow\infty} r^2\,\hat{x}_i\,\omega_i(x)^\pm\neq 0.
\end{eqnarray}
One such typical $\omega_i^+$ has the Fourier transform
\begin{eqnarray}
\label{omega}
\hat{\omega}_i^+(k)=\int d^3x\, e^{i\vec{k}\cdot\vec{x}}\omega_i^+(x)=\frac{1}{P\cdot k+i \epsilon}
(P_i-\vec{P}\cdot\hat{k} \,\hat{k}_i)
\end{eqnarray}
(with  $\epsilon$ decreasing to zero as usual).
The momentum $P_\mu$ is the total momentum of the charged system. (We have not shown the individual momenta and charges
of which $P$ and $q_n$ are composed as they are not importatnt for our considerations.) The important point here is that
$\hat{\omega}_i^+$ is
not square-integrable:
\begin{eqnarray}
\langle \omega,\omega\rangle:=\lim_{|\vec{k}|\rightarrow 0}\int^\infty_{|\vec{k}|}
\frac{d^3k}{2|\vec{k}|}|\hat{\omega}_i^+(k)|^2=\infty.
\end{eqnarray}
It is then a theorem~\cite{ref3} that the representation of $\mathcal{A}$ built on (\ref{reprA})
is superselected: it is not the Fock space representation.

The appendix gives a derivation of (\ref{reprA}). 

We now elaborate on the physical meaning of  (\ref{reprA}).  Consider the current
\begin{eqnarray}
\label{Jcurrent}
J^\mu(x)=q_n\int d\tau \,\delta^4(x-z(\tau))\frac{dz^\mu(\tau)}{d\tau}.
\end{eqnarray}
It radiates photons of momenta $k:=(|\vec{k}|,\vec{k})$. We are interested in infrared photons, so we assume that
\begin{eqnarray}
\frac{dz^\mu(\tau)}{d\tau}=\frac{P^\mu}{m}, \quad\quad
z^{\mu}(\tau)=\tau \frac{P^\mu}{m}, \quad\quad m^2=P^\mu P_\mu, \quad\quad m>0,
\end{eqnarray}
where $P^\mu$ is constant.

Now (\ref{Jcurrent}) generates the additional interaction 
\begin{eqnarray}
\int d^3 x A_\mu(x) J^\mu(x). 
\end{eqnarray}
It changes the ``in" state to  $|n, \omega \cdots \rangle$ as is shown in the appendix.

The following is a further important point. If the Lorentz boost
\begin{eqnarray}
K_i=\int d^3x\, x_i [\vec{E}^2(x)+\vec{B}^2(x)]+\textrm{matter part}
\end{eqnarray}
is well-defined in $|n,0;\cdot\rangle$, then it diverges in the sector $|n,\omega;\cdot\rangle$, $\omega\neq 0$ :
Lorentz invariance is broken in the latter. We can see this as follows:
\begin{eqnarray}
&&\langle n,\omega;\cdot|\int d^3x \, x_i[\vec{E}^2(x)+\vec{B}^2(x)]|n,\omega;\cdot\rangle \nonumber \\&&=
\langle n,0;\cdot|\int d^3x\, x_i [ (\vec{E}-\vec{\omega})^2(x)+\vec{B}^2(x)]|n,0;\cdot\rangle,\nonumber \\
&& \omega_i := \omega_i^+ + \omega_i^-
\end{eqnarray}
and this diverges since $\vec{\omega}^2(x)=\mathcal{O}(\frac{1}{|\vec{x}|^4})$ as $|\vec{x}|\rightarrow\infty$.

There is an alternative approach for these considerations due to Roepstorff~\cite{ref3}. It is based
on the Weyl algebra and the GNS construction. For free scalar fields, in four-dimensional spacetime, the
Weyl algebra $\mathcal{W}$ has elements $W(f)$ where $f$ is a compactly supported real test function, $f\in
\mathcal{C}_0^\infty$. If $g$ is another such test function,
\begin{eqnarray}
&& W(f)W(g)=W(f+g)e^{i \sigma(f,g)/2}, \\
&& \sigma(f,g)=i\int d^4x\, d^4y\, f(x) D(x-y) g(y), \\
&& D(x-y)=\textrm{commutator function}=\int \frac{d^3P}{(2\pi)^3}\frac{1}{2|P_0|}[
e^{-iP\cdot(x-y)}-e^{iP\cdot(x-y)}].
\end{eqnarray}
Since $(\Box+m^2)D=0$ if the field has mass $m$, $\sigma$ vanishes on any function $f$ of the form $(\Box+m^2)h$, with
$h\in\mathcal{C}_0^\infty$. On quotenting out such functions, $\sigma$ becomes a symplectic form.

Also, the function $\sigma(f,\cdot)$ defined by
\begin{eqnarray}
\sigma(f,\cdot)(y)=i\int d^4x\, f(x) D(x-y)
\end{eqnarray}
fulfills the equations of motion.

Let us introduce the scalar product
\begin{eqnarray}
(f,g)=\frac{1}{(2\pi)^3}\int \frac{d^3k}{2|k_0|}\overline{\tilde{f}(k)}\tilde{g}(k), \quad\quad
|k_0|=(\vec{k}^2+m^2)^{1/2}, \nonumber
\end{eqnarray}
where
\begin{eqnarray}
\tilde{f}(k)=\int d^4x\, e^{-i k\cdot x}f(x), \quad\quad \tilde{g}(k)=\int d^4x\, e^{-i k\cdot x} g(x).
\end{eqnarray}
Then the Fock representation with the corresponding Hilbert space $\mathcal{H}$ is given by the following state $\omega_0$ on the Weyl algebra and the GNS construction:
\begin{eqnarray}
\omega_0(W(f))=e^{-(f,f)/2}.
\end{eqnarray}
If $|0\rangle$ is the Fock vacuum, we have, as can be checked,
\begin{eqnarray}
\omega_0(W(f))=\langle 0|W(f)|0\rangle.
\end{eqnarray}

Now if $F$ is a linear functional on test functions, we can twist $\omega_0$ to a new state $\omega_F$:
\begin{eqnarray}
\omega_F(W(f))=\omega_0(W(f))e^{i\,\textrm{Im} \,F(f)}.
\end{eqnarray}
Suppose we can write
\begin{eqnarray}
F(f)=\langle \eta,f\rangle, \quad\quad \eta\in\mathcal{H}.
\end{eqnarray}
Then, by Schwarz inequality,
\begin{eqnarray}
\| F(f)\|\leq \|\eta\|\|f\|.
\end{eqnarray}
Conversely, by the Riesz-Frechet theorem~\cite{ref9},
we can write $F(f)=\langle\eta,f\rangle$ for $\eta\in\mathcal{H}$
iff
\begin{eqnarray}
\|F(f)\|<c\|f\|, \quad\quad c\,\,  =\textrm{ constant}.
\end{eqnarray}

Now if there is such an $\eta$, we can check that
\begin{eqnarray}
\omega_F(W(f))=\langle 0| W(\textrm{Im}\,\eta)^\dagger W(f) W(\textrm{Im}\,\eta)|0\rangle.
\end{eqnarray}
Since $W(\textrm{Im}\,\eta)|0\rangle$ is in the Fock space, the GNS representation from $\omega_F$ is unitarily equivalent to
the one from $\omega_0$.

Any smooth $\xi\in\mathcal{C}^\infty$ which is not in $\mathcal{H}$ also gives an $F$:
\begin{eqnarray}
F(f)=(\xi,f),
\end{eqnarray}
since $f$ is a test function and hence compactly supported. In this case,
\begin{eqnarray}
\omega_F(W(f))=\langle0|W(\textrm{Im}\,\xi)^\dagger W(f)W(\textrm{Im}\,\xi)|0\rangle,
\end{eqnarray}
but $W(\textrm{Im}\,\xi)|0\rangle$ is not in the Fock space. So the representation of $\mathcal{W}$ built on $W(\textrm{Im}\,\xi)|0\rangle$
is not unitarily equivalent to the Fock representation.

These considerations can be adapted to QED. Let $\alpha_\mu$ be the test function for the potential in the Lorentz
gauge $\partial^\mu\alpha_\mu=0$. If
\begin{eqnarray}
f_{\mu\nu}(\alpha)=\partial_\mu\alpha_\nu-\partial_\nu\alpha_\mu,
\end{eqnarray}
the symplectic form (modulo the kernel) is $\sigma$:
\begin{eqnarray}
\sigma(\alpha_1,\alpha_2)=i\int d^4x \,d^4y\, \alpha_1^\mu(x)D(x-y)\alpha_{2\mu}(y).
\end{eqnarray}
It depends only on $f_{i,\mu\nu}=\partial_\mu \alpha_{i\nu}-\partial_\nu\alpha_{i,\mu}$, as it is gauge invariant:
$\sigma(\alpha_1,\alpha_2)$ remains invariant under $\alpha_\mu\rightarrow \alpha_\mu+\partial_\mu\eta$. Thus, the Weyl
algebra is defined by
\begin{eqnarray}
W(\alpha_1)W(\alpha_2)=W((\alpha_1+\alpha_2))e^{\frac{i}{2}\sigma(\alpha_1,\alpha_2)}.
\end{eqnarray}
With
\begin{eqnarray}
&&\tilde{\alpha}_\mu(k)=\int d^4x\, \alpha_\mu(x)e^{-ik\cdot x}, \quad\quad k_0=|\vec{k}|, \\
&& \implies k^\mu\tilde{\alpha}_\mu(k)=0,
\end{eqnarray}
we introduce the scalar product
\begin{eqnarray}
(\alpha_1,\alpha_2)=\frac{1}{(2\pi)^3}\int \frac{d^3k}{2|\vec{k}|}\sum_{i=1}^3(\overline{\tilde{\alpha}}_{1,i}-\frac{k_i}{k_0}
\overline{\tilde{\alpha}}_{1,0})(\tilde{\alpha}_{2,i}-\frac{k_i}{k_0}\tilde{\alpha}_{2,0})(k).
\end{eqnarray}
We can then adapt the scalar field considerations to the electromagnetic field.

The choice
\begin{eqnarray}
F(\alpha)=(\omega,\alpha)
\end{eqnarray}
when $\omega$ is given by (\ref{omega}) leads to the vertex operator in (\ref{reprA}).

We note that 
 \begin{eqnarray}
 \label{Falpha}
	F(\alpha)=  \int dx \,dy\, J^\mu (x)D (x-y) \alpha_\mu (y)
\end{eqnarray}
if the $\tau$ integration in $J_\mu$ is restricted to $\tau\leq0$ and the Lorentz gauge is changed to the Coulomb gauge.
See Appendix.

\section{The Case of QCD}

\label{section3}

By the axioms of QFT~\cite{ref9}, observables are local. They generate the algebra $\mathcal{A}$ of local
observables.

The group $SU(3)$ of QCD commutes with all elements of $\mathcal{A}$, just as the $U(1)$ of charge commutes with
all elements of $\mathcal{A}$. But $SU(3)$ unlike $U(1)$ is non-abelian. For this reason, we
argued~\cite{ref1,ref7} that $SU(3)$ is spontaneously broken.

Our arguments were as follows. If $\mathbb{C}SU(3)$ is the group algebra of $SU(3)$, $\mathbb{C}SU(3)$ also
commutes with $\mathcal{A}$. A maximal abelian subalgebra of $\mathbb{C}SU(3)$  is spanned by the complete
commuting set (CCS) $c_2,c_3,\hat{\lambda}_3,\hat{\lambda}_8$ where $c_2,c_3$ are the quadratic and cubic
Casimir operators and $\hat{\lambda}_3,\hat{\lambda}_8$ are the operators which represent the $\lambda_3,\lambda_8$
of the Gell-Mann matrices on the Hilbert space. In any IRR of $\mathcal{A}$, we can diagonalise the CCS.
The basis vectors in this
IRR can be written as
\begin{eqnarray}
|c_2,c_3,i_3,y;\cdot\rangle,
\end{eqnarray}
where $c_2,c_3,i_3,y$ are the eigenvalues of $c_2,c_3,\hat{\lambda}_3,\hat{\lambda}_8$. No observable can affect
them. They label a superselection sector.

But in a coloured representation $(c_2,c_3 \textrm{ or both}\neq0)$, a generic $SU(3)$ transformation will
change $i_3,y$. That is, it will change the superselection sector. Hence, such an $SU(3)$ transformation is spontaneously broken.

We have analysed a similar situation which occurs for the ethylene molecule~\cite{ref10}. The gauge group there is
$D_8^\ast$, the binary dihedral group isomorphic to the group 
\begin{eqnarray}
\{\pm\mathbb{I}_{2\times 2}, \pm i\tau_j: \quad j=1,2,3, \quad \tau_j=\textrm{ Pauli matrices}\}.
\end{eqnarray}
This is non-abelian. Its two-dimensional representation is relevant for certain confirmations of the molecule.
We proved explicitly that its maximal abelian subgroup, say
\begin{eqnarray}
H=\{\pm\mathbb{I}_{2\times 2},\pm i\tau_3\}
\end{eqnarray}
is diagonal in an IRR of the molecule: no observable changes the eigenvalues of elements of $H$.

A consequence of the above observation is that coloured states, such as
\begin{eqnarray}
\label{colstate}
|c_2,c_3,i_3,y;\cdot\rangle\langle c_2,c_3,i_3,y;\cdot|
\end{eqnarray}
for $c_2,c_3 \textrm{ or both}\neq0$, are {\it mixed} \cite{ref7}. That has consequences for colour
confinement~\cite{ref7}.

In what follows, we give another argument to show that  the $SU(3)$ of colour is spontaneously broken. It uses the
infrared cloud in QCD which dresses the coloured particle. An extra result we find is that Lorentz invariance
is broken in QCD too.

We approach this problem as in the QED case. The charged particle in QED gets dressed by the infrared radiation and
becomes the in and out state. The latter is not in the Fock space and breaks Lorentz invariance.

In a similar manner, we can expect novel physical consequences from the gluon radiation of Yang-Mills particles
such as quarks.

The dynamics of Yang-Mills gluons is non-linear. Hence, gluons can also radiate gluons. We do not treat
this radiation. Perhaps by using a null four-vector for $dz_\mu(\tau)/d\tau$ in what follows we can get
an adequate description of infra-gluon radiation by gluons.

Nor do we treat effects of confinement.

The electric charge $e$ is replaced by an ``internal'' vector of operators $\hat{\lambda}=
(\hat{\lambda}_\alpha, \,\,\alpha=1,2,\ldots,8)$ for a Yang-Mills or Wong~\cite{ref11} particle. (We absorb the Yang-Mills coupling constant in the gluon field.)
The operators $\hat{\lambda}$ are $SU(3)$ Lie algebra valued with the standard commutators
\begin{eqnarray}
\label{comm}
[\hat{\lambda}_\alpha,\hat{\lambda}_\beta]=if_{\alpha\beta}^{{\color{white}\alpha\beta}\gamma} \hat{\lambda}_\gamma.
\end{eqnarray}
Their representation on the quantum Hilbert space identifies the particle as a quark, a di-quark, etc.

The quantum Yang-Mills current is
\begin{eqnarray}
J_\mu^\alpha(x)=\int d\tau \,\delta^4(x-z(\tau))\hat{\lambda}^\alpha(\tau)\frac{dz_\mu(\tau)}{d\tau}.
\end{eqnarray}
The vector $\hat{\lambda}(\tau)$ can change with $\tau$, fulfilling (\ref{comm}).

There is a consistency condition on $J_\mu$: it must be covariantly constant. Thus the field equation gives
\begin{eqnarray}
\label{eqn2}
D^\mu F_{\mu\nu}=J_\nu, \quad\quad F=\textrm{ curvature of the gluon connection } A,
\end{eqnarray}
where
\begin{eqnarray}
D^\mu D^\nu F_{\mu\nu}=0
\end{eqnarray}
as an identity. Hence,
\begin{eqnarray}
D^\mu J_\mu =0
\end{eqnarray}
or
\begin{eqnarray}
\label{Dlambda}
D_\tau\hat{\lambda}(\tau)\equiv \,\,\, \partial_\tau\hat{\lambda}(\tau) +i\frac{dz^\mu(\tau)}{d\tau}[A_\mu(z(\tau)),\hat{\lambda}(\tau)]=0.
\end{eqnarray}
Thus if
\begin{eqnarray}
&& g(\tau)=P \textrm{exp}\left(-i\int_0^\tau d\tau^\prime\frac{dz^\mu(\tau^\prime)}{d\tau^\prime}A_\mu
(z(\tau^\prime))\right), \\
&& \implies [\partial_\tau+iA_\tau(z(\tau))]g(\tau)=0,
\end{eqnarray}
then
\begin{eqnarray}
\label{transf}
\hat{\lambda}(\tau)=g(\tau)\hat{\lambda}(0)g(\tau)^{-1}.
\end{eqnarray}
This equation determines the evolution of $\hat{\lambda}$ with $\tau$. Since $g(\tau)\in SU(3)$, this is by a gauge
transformation.

Because of (\ref{transf}), the $\tau$-evolution on $J$ does not mix different IRR's of $\hat{\lambda}$ and
commutes with the Casimir invariants. We therefore assume that $\hat{\lambda}$ belongs to a fixed IRR
of $\underline{SU(3)}$ (the underline denoting Lie algebra).

The Wong equations have a Lagrangian description found by us long ago \cite{ref12}.

\subsection{The Gauge Choice}

Let $K$ be a fixed element of $\underline{SU(3)}$ which is in the orbit of $\hat{\lambda}(0)$. Then we can
find $h(\tau)\in SU(3)$ such that
\begin{eqnarray}
K=h(\tau)\hat{\lambda}(\tau)h(\tau)^{-1}.
\end{eqnarray}
That is because the $\tau$-evolution (\ref{transf}) is by gauge transformation.

Let us assume that $K$ is a ``generic'' element of $\underline{SU(3)}$, so that its stability
group $S[U(1)\times U(1)]$ is abelian. (We comment on non-abelian stability groups below.)
In that case,
\begin{eqnarray}
A_\tau[z(\tau)]:=\frac{dz^\mu(\tau)}{d\tau} A_\mu[z(\tau)]\in \underline{U(1)\times U(1)}.
\end{eqnarray}
That is because $A_\tau[z(\tau)]$ commutes with $K$ by (\ref{Dlambda}).

The current is now abelian, in the direction $K$:
\begin{eqnarray}
J^\mu(x)=K\int d\tau \delta^4[x-z(\tau)]\frac{dz^\mu(\tau)}{d\tau}.
\end{eqnarray}

We must now solve (\ref{eqn2}) for the emitted radiation at low frequencies by this current. That can be done as in
Peshkin and Schr\"oeder \cite{ref8}, by setting
\begin{eqnarray}
A_\mu=a_\mu K+b_\mu L,
\end{eqnarray}
where $\textrm{Tr}(KL)=0$, $K,L$ are the basis of the two $\underline{U(1)}$'s and commute, and $a_\mu(x),b_\mu(x)\in\mathbb{R}$.
Then, from (\ref{eqn2}),
\begin{eqnarray}
\partial^\mu(\partial_\mu a_\nu-\partial_\nu a_\mu)K=J^\mu,
\end{eqnarray}
whereas $b$ has no source.

This equation for $a$ can be treated exactly as in QED. The particle state vectors
have eigenvalues $k_i$ of $K$. So the radiation is $k_j\omega_\mu$ for a fixed $k_j$ and its dressed in state is
\begin{eqnarray}
\label{eqn}
|\textrm{in},\omega,\ldots\rangle=e^{k_j\int d^3x(\textrm{Tr} \lambda_K [A_i^-(x)\omega_i^+(x)-A_i^+(x)\omega_i^-(x)]}
|P;k_j\rangle
|0\rangle_\gamma,
\end{eqnarray}
where $\lambda_K= K/{\sqrt{\textrm{Tr} K^2}}$, $|0\rangle_\gamma$ is the gluon Fock vacuum and $|P;k_j\rangle$ is
the particle  state with eigenvalue
$k_j$ for $K$.

In~\cite{ref7}, we have shown how the twist of the in state vector in (\ref{eqn}) can be absorbed as an instanton-like
term in the action.

\subsection{Lorentz Invariance}

Just as in QED, Lorentz invariance is spontaneously broken in QCD. The proof follows from (\ref{eqn}) as in QED.

\section{QCD Breaks Colour}

\label{section4}

The considerations leading to the gauge choice $K$ imply that the test functions $\alpha$ are proportional to
$K:\,\alpha=K\beta$, where $\beta$ are real-valued. The linear functional $F$ in (\ref{Falpha}) now involves a trace over the
Lie algebra elements.
Correspondingly, there is now a trace in (\ref{Falpha}). Also, the electric charge $q_n$ is replaced by $k_j$.

Now $K$ has in general many eigenvalues and we can transform one eigenvalue $k_j$ to another, $k_l$,  by conjugating $K$ by a suitable $SU(3)$
element. Consider the case where $k_j\neq k_l$. The change in $F(\alpha)$ is then by
\begin{eqnarray}
   (\zeta ,\alpha)  
\end{eqnarray}
where
\begin{eqnarray}
\zeta^\mu = (k_l-k_j)\int_{\tau\leq0} d^4y\, d\tau\, \delta^4(y-z(\tau)) \frac{dz^\mu}{d\tau}  D(y-x).
\end{eqnarray}
But $\zeta$ is not square integrable and $(\zeta,f)$ does not fulfill the bound $| (\zeta,f)| \leq c||f||$,
with $c=\textrm{constant}$.
Thus by the  theorem in~\cite{ref3}, this transformation cannot be unitarily implemented.
  Thus the gauge transformation on $K$ which changes its eigenvalue from $k_j$ to $k_l$
  cannot be unitarily implemented. That is, $SU(3)$ is spontaneously broken.

  The change from $k_j$ to $k_l \neq  k_j$ is very much like changing the charge in QED from $q_n$ to
  $q_m\neq q_n$.
  We see that just as in the QED  case, such a change is not unitarily implementable.
  
  {\it Remarks on phenomenogy:} After the gauge choice which leads to (\ref{eqn}), the vertex operator there is as in QED~\cite{ref7},
  with $k_j$ replacing the total electric charge. Therefore, we can twist the mass term of coloured particles as in that paper. In QCD as well,
  that explicitly realises the known QED results~\cite{ref4,ref5} about the  mass and spin shells of particles not being sharp.
  It also gives simple models for phenomenolgy and shows, for example, that life times and decays of coloured particles are frame dependent.

\subsection{The Case where $K$ is Non-Generic}

In this case, $K$ is parallel to $\hat{\lambda}_8$ and its stability group is non-abelian,
being $S(U(2)\times U(1))$, which is isomorphic  to $U(2)$. Its Lie algebra is the direct sum of those of $SU(2)$
and $U(1)$, where the latter has $\lambda_8$ as a basis. Also, $\underline{U(1)}$ commutes with $\underline{SU(2)}$. 
The generators of $\underline{SU(2)}$ and $\underline{U(1)}$ are trace orthogonal while the current $J_\mu$
has only the $\underline{U(1)}$ or $\lambda_8$ component. Thus the current is non-zero only for the abelian component
$\lambda_8$
of $A_\tau(z(\tau))$. The $\underline{SU(2)}$ part of $A_\tau(z(\tau))$ decouples from this $\underline{U(1)}$ part.
So the $\underline{U(1)}$ part can be treated as before, with similar conclusions.

\section*{Acknowledgements}

The original claim that non-abelian superselection rules are spontaneously broken is contained in the paper with Vaidya~\cite{ref1}.
In papers with Queiroz and Vaidya~\cite{ref10,ref13}, the related result was shown that coloured states are mixed and hence cannot
be produced from colour singlets. The work on Lorentz symmetry breaking was done with K\"urk\c{c}\"uo\u{g}lu,
Queiroz and Vaidya~\cite{ref7}.
I am grateful to all these colleagues and to Kumar Gupta  for valuable inputs.
I   very much appreciate the invaluable help I received from Nirmalendu Acharyya and Ver\'onica Errasti D\'iez
in preparing the paper. Vero was especially helpful, carefully reading the paper and suggesting improvements.
I also thank Andr\'es Reyes at the Universidad de Los
Andes for hosting me as the Sanford Professor and for warm hospitality while this work was being completed.

\appendix
\section{Appendix: Derivation of (\ref{reprA}) }
This appendix gives a derivation of equation (\ref{reprA}). It is not new and its variations may be found in several
papers( cf. \cite{Eriksson:1970dc, ref3, ref5}).  We describe it here to help the reader.

Equation (\ref{reprA}) shows  the in-state of the charged particle of momentum $P$ and charge $q_n$ after it is dressed by
the cloud of infrared photons. If there are several particles with total  momentum $P$ and charge $q_n$, we can
approximate them by a single particle of momentum $P$ and charge $q_n$ as explained by previous works.

We start with non-interacting charges and photons, the latter being in their vacuum state: 

\begin{eqnarray}\label{app1}
|n, P, \cdots \rangle \otimes |0\rangle_\gamma.
\end{eqnarray}
Here $|0\rangle_\gamma$ is the photon vacuum amd $n$ as before denotes charge $q_n$. Let $H_0$ be the Hamiltonian associated with (\ref{app1}).

The current $J^\mu$ of the charged particle can be approximated by that of a particle moving with charge $q_n$ and
constant velocity $\frac{P^\mu}{(P \cdot P)^{\frac{1}{2}}}:= \frac{P^\mu}{M}$. It radiates photons of various momenta,
but since we are interested in the infrared limit where they approach zero, we can assume that $\frac{P^\mu}{M}$ is constant.

The current $J^\mu$ is thus 
\begin{eqnarray}
J^\mu (x) &=& q_n \int_{-\infty}^{\infty} d \tau \,\, \delta^4 (x - z(\tau)) \frac{ d z^\mu(\tau)}{d \tau} \nonumber \\
&=& \int_{-\infty}^{0} d \tau \,\,  \delta^4 (x - \tau \frac{P}{M}) \frac{P^\mu}{M}
\end{eqnarray}
where for the charged particle tragectory $z(\tau)$, we assume that 
\begin{equation}
z^\mu (\tau) = \tau \frac{P^\mu}{M}.
\end{equation}

The interaction of $A_\mu$ with $J^\mu$ is described by the interaction Hamiltonian 
\begin{eqnarray}
H_I (t) &=& q_n \int d^3 x A_\mu (x) \int d \tau \frac{P^\mu}{M}\,\,  \delta^4(x - \tau \frac{P}{M}) \\
&=&q_n \left(\frac{P^\mu}{P^0} \right) A_\mu \left(x_0\frac{P}{P^0} \right), \quad\quad t=x_0. 
\end{eqnarray}

The in-state associated with (\ref{app1}) is obtained by applying the M{\o}ller operator
\begin{eqnarray}
\Omega = \lim_{t \rightarrow -\infty} e^{i H t} e^{ - i H_0 t} \equiv \lim_{t \rightarrow -\infty}  U(t)
\end{eqnarray}
to (\ref{app1}). It is  (\ref{reprA}) where $\omega$ will make its appearance below. Thus
\begin{eqnarray}
|n, P, \omega, \cdots\rangle &=& \Omega | n, P, \cdots \rangle \otimes |0\rangle_\gamma \nonumber \\
&\equiv& \Omega |n, P, 0,\cdots\rangle \otimes  |0 \rangle_\gamma
\end{eqnarray}

Now 
\begin{eqnarray}
i \frac{d U(t)}{dt} &=& - U(t) \hat{H}_I(t) \nonumber \\
 \hat{H}_I(t)  &=& e^{i H_0 t} H_I (t) e^{- i H_0 t},
\end{eqnarray}
$ \hat{H}_I(t) $ being the interaction Hamiltonian in the interaction representation.

Thus
\begin{eqnarray}\label{app9}
U(t) = T\,\, exp \left[ -i \int_t^0 dt^\prime \hat{H}_I (t^\prime) \right] 
\end{eqnarray}
if $\hat{A}_\mu$ is the vector potential in interaction representation, we can also write (\ref{app9})  as a $T$-ordered Dirac-Wilson line integral
\begin{eqnarray}
U(t) = T\,\, exp \left[ -i \int_t^0 dz^\mu(x_0) A_\mu\left(x_0 \frac{P}{M}\right)\right].
\end{eqnarray}

Since
\begin{eqnarray}
[[\hat{H}_I (t),\hat{H}_I(t^\prime)],\hat{H}_I(t^{\prime\prime}] =0
\end{eqnarray}
it is known that \cite{Eriksson:1970dc}
\begin{eqnarray}
U(t) = exp \left[ -i \int_t^0 dt^\prime \hat{H}_I (t^\prime) \right] \times u(t)
\end{eqnarray}
where $u(t)$ is a (time- and $P$-dependent) phase commuting with all operators of the electromagnetic field. It does not affect us. Hence we can drop it from $\Omega$ and write
\begin{eqnarray}
\Omega &=& exp \left[ -i \int_{-\infty}^0 dt^\prime \hat{H}_I (t^\prime) \right] \\
&=&exp \left[ -i q_n\int_{-\infty}^0 dz^\mu(x_0) A_\mu\left(x_0 \frac{P}{M}\right)\right]
\end{eqnarray}

Let 
\begin{eqnarray}
 A_\mu^+(x) = \int \frac{ d^3 k}{\sqrt{2 k}}\frac{1}{ (2 \pi)^3} a_\mu^+(k) e^{i k \cdot x}, \quad\quad k \cdot x = k_0 x_0 - \vec{k} \cdot \vec{x} 
\end{eqnarray}
denote the positive frequency part of $A_\mu(x)$ and let 
\begin{eqnarray}
A_\mu^- (x) = (A_\mu^+(x))^\dagger, \quad\quad a_\mu^- (k) = (a_\mu^+(k))^\dagger.
\end{eqnarray}

Then
\begin{eqnarray}
&& q_n\int_{-\infty}^0 dx_0 \frac{P^\mu}{P^0} A_\mu (x_0 \frac{P}{P^0}) = \nonumber \\
&& - i \int \frac{ d^3 k}{\sqrt{2 k_0}}\frac{1}{ (2 \pi)^3} a_\mu^+(k) \tilde{\omega}^{\mu -}(k)+ i  \int \frac{ d^3 k}{\sqrt{2 k_0}}\frac{1}{ (2 \pi)^3} a_\mu^-(k) \tilde{\omega}^{\mu +}(k), \\
&&  \tilde{\omega}^{\mu +} (k) = \frac{ q_n P^\mu}{k \cdot P + i \epsilon} =  \bar{\tilde{\omega}}^{\mu +} (k)
\end{eqnarray}
where $\epsilon$ as usual decreases to zero when all calculations are done.

Thus we get
\begin{eqnarray}\label{app19}
\Omega = exp -\left\{\int \frac{ d^3 k}{\sqrt{2 k_0}}\frac{1}{ (2 \pi)^3} [a_\mu^+(k) \tilde{\omega}^{\mu -}(k)-a_\mu^-(k)
\tilde{\omega}^{\mu +}(k)]\right\}. 
\end{eqnarray}

The gauge choice in (\ref{app19}) is Lorentz: $k^\mu a_\mu^\pm (k)=0$. So we can replace $a_\mu^\pm (k)$ by 
\begin{eqnarray}
a_\mu ^{\pm \,\,  \prime} (k) = a_\mu^\pm (k) - \frac{k_\mu}{k_0} a_0^\pm (k) 
\end{eqnarray}
which is in the Coulomb gauge: 
\begin{equation}\label{app21}
a_0^{\pm \,\, \prime} (k) =0, \quad\quad k_i a_i^{\pm \,\, \prime}(k) =0. 
\end{equation}

Hence
\begin{equation}
a_\mu^\mp (k) \tilde{\omega}^{\mu \,\, \pm} \rightarrow a_i^{\mp  \,\, \prime} (k) \tilde{\omega}^{i  \,\, \pm}.
\end{equation}
Since $a_i^{\pm  \,\, \prime} $ are transverse by (\ref{app21}), we can now change $\tilde{\omega}^{i \pm}$ to 
\begin{eqnarray}
\hat{\omega}^{i  \,\, \pm} (k) = \tilde{\omega}^{i \,\,  \pm} (k) - \hat{k}^i \hat{k}^j \cdot \tilde{\omega}^{j  \,\, \pm}(k).
\end{eqnarray}

On Fourier transforming 
\begin{eqnarray}
\int \frac{ d^3 k}{\sqrt{2 k_0}}\frac{1}{ (2 \pi)^3} [a_\mu^{- \,\,  \prime}(k) \hat{\omega}^{\mu \,\,  +}(k)-a_\mu^{+ \,\,  \prime}(k) \hat{\omega}^{\mu \,\,  -}(k)]
\end{eqnarray}
and calling 
\begin{eqnarray}
\int \frac{ d^3 k}{\sqrt{2 k_0}}\frac{1}{ (2 \pi)^3} [a_\mu^{+ \,\,  \prime}(k)e^{ik \cdot x}+a_\mu^{- \,\,  \prime}(k) e^{-ik \cdot x}]
\end{eqnarray}
as $A_i (x)$ (with no prime), we get  (\ref{reprA}). This $A_i$ is in Coulomb gauge.

\end{document}